\documentclass[5p]{elsarticle}

\usepackage{lineno}
\usepackage{booktabs }
\usepackage[colorlinks,linkcolor=blue]{hyperref}
\usepackage{amsmath}
\usepackage{epstopdf}
\modulolinenumbers[5]

\journal{Journal of \LaTeX\ Templates}

\begin{document}

\begin{frontmatter}

\title{Longitudinal RF capture simulation and BPM signal estimation}
\author[mymainaddress,mysecondaryaddress]{Yongchun Feng}
\author[mymainaddress]{Ruishi Mao\corref{mycorrespondingauthor}}
\cortext[mycorrespondingauthor]{Corresponding author}
\ead{maorsh@impcas.ac.cn}
\author[mymainaddress]{Yucong Chen}
\author[mymainaddress]{Yan Yin}
\author[mymainaddress]{Xiaohu Zhang}
\author[mymainaddress,mysecondaryaddress]{Shuang Ruan}
\author[mymainaddress,mysecondaryaddress]{Tong Liu}
\author[mymainaddress,mysecondaryaddress]{Yaoyao You}
\author[mymainaddress]{Xincai Kang}
\author[mymainaddress]{Tiecheng Zhao}
\author[mymainaddress]{Zhiguo Xu}
\author[mymainaddress]{Peng Li}
\author[mymainaddress]{Yanyu Wang}
\author[mymainaddress]{Youjin Yuan}

\address[mymainaddress]{Institute of Modern Physics, Chinese Academy of Science, Lanzhou 730000, China}
\address[mysecondaryaddress]{University of Chinese Academy of Sciences, Beijing 100049, China}

\begin{abstract}
In this paper, the theoretical aspects behind longitudinal RF capture are reviewed and the capture process is simulated via a program based on this theory. Four kinds of cases with different initial distribution and capture curve are considered, i.e. uniform distribution with adiabatic capture, uniform distribution with non-adiabatic capture, Gaussian distribution with adiabatic capture and Gaussian distribution with non-adiabatic capture. The simulation results are compared each other and discussed, and Gaussian distribution with adiabatic capture is demonstrated having a higher capture efficiency and leading to a shorter bunch length. In addition, the BPM induced signal is simulated with high input impendence, i.e. $1M\Omega$, and low input impendence, i.e. $50\Omega$, respectively. Finally, the BPM signal of Heavy Ion Medical Machine (HIMM) is estimated and compared with measured one, and a good agreement is achieved.
\end{abstract}


\end{frontmatter}

\linenumbers
\section{\label{intro}Introduction}
BPM is used in many accelerator facilities to control beam orbit. There exists various types of BPMs, such as button BPM for electron accelerators and proton linac, linear cut BPM for proton and heavy ion synchrotrons, and stripline BPM for short bunch observation. However, bunched beam is essential for a BPM to induce a significant signal. For most synchrotrons, the injected beam is always stored as a coasting beam after many revolutions due to debunching. So, a RF cavity is indispensable to capture coasting beam to a bunched one for BPM observation. Meanwhile, in order to accelerate beam to a high energy, the coasting beam has to be captured with a high capture efficiency to perform a high acceleration efficiency. In general, capture process can be considered as adiabatic and nonadiabatic. Adiabatic capture states that parameters in the synchrotron Hamiltonian change slowly so that the particle orbit is a torus of constant Hamiltonian value\cite{Lee1}, and verse vice for nonadiabatic capture. To obtain high capture efficiency, adiabatic capture should be chosen firstly, however, which leads to a long capture time that is not suitable for rapid cycling synchrotron. So, in some sense, a trade-off has to be made between high capture efficiency and short capture time. In this article, two kinds of capture curves corresponding to adiabatic and nonadiabatic capture respectively are investigated and a comparison is discussed.
\begin{figure}
\includegraphics[width=8cm]{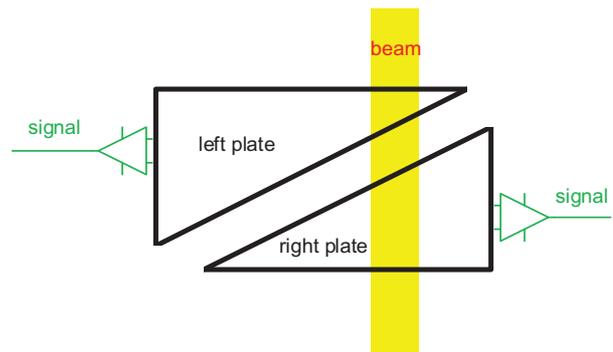}
\caption{\label{fig1}   Schematic of a typical linear cut BPM.}
\end{figure}

BPM is of great importance for accelerator commission and operation, which consists of two pairs of plates inserted in the beam pipe, and horizontal and vertical beam center are determined by the induced signal on the plates. While many kinds of BPMs there exist, we limit our analysis to capacitive linear cut BPM, or shoe box BPM, which is widely used in proton and heavy ion synchrotrons such as HIMM. Although we considered only linear cut BPM here, the signal treatment methods we used can be applied to any types of capacitive BPMs. The schematic of a typical linear cut BPM is shown in Fig.\ref{fig1}. Such type of BPM can induce a precise linear dependence signal with respect to beam displacement, that is to say, having a good linearity. In this article, BPM signal is estimated for different input impendence, and an example is taken to deepen the understand of BPM signal.
\section{The synchrotron mapping equation and the stability of algorithm}
In the synchrotron phase space coordinates $(\phi,\Delta E)$, the mapping equations can be written as\cite{Lee1}
\begin{equation}
\label{eq1}
\left\{
\begin{aligned}
\Delta E_{n+1} & =  \Delta E_{n}+qV(sin\phi_n-sin\phi_s) \\
\phi_{n+1} & =  \phi_{n}+\frac{2\pi h\eta}{\beta^2E}\Delta E_{n+1}
\end{aligned}
\right.
\end{equation}
in which $\Delta E=E-E_s$, $\phi$ and $E$ are respectively the synchronous phase angle and the total energy for an off-momentum particle. $E_s$ refers to the total energy for an on-momentum particle. $q$ is the particle charge, $V$ is the RF voltage amplitude, $h$ is the harmonic number, $\eta=\frac{1}{\gamma^2}-\frac{1}{\gamma^2_T}$ is the phase slip factor, $\gamma=\frac{1}{\sqrt{1-\beta^2}}$ is the relativistic factor and $\gamma_T$ is the transition energy factor.

To check whether the algorithm is stable or not, the Jacobian for variables transformation plays an important role. The phase space area from $(\Delta E_{n}, \phi_{n})$ to $(\Delta E_{n+1}, \phi_{n+1})$ can be formulated as
\begin{equation}
\label{eq2}
\iint d\Delta E_{n+1}d\phi_{n+1}=\iint J(\Delta E,\phi)d\Delta E_{n}d\phi_{n}
\end{equation}
where the Jacobian is
\begin{equation}
\label{eq3}
\begin{aligned}
J(\Delta E,\phi)&=
\left|
\begin{array}{ccc}
\frac{\partial \Delta E_{n+1}}{\partial \Delta E_{n}} & \frac{\partial \Delta E_{n+1}}{\partial\phi_{n}} \\
\frac{\partial \phi_{n+1}}{\partial \Delta E_{n}} & \frac{\partial \phi_{n+1}}{\partial\phi_{n}}
\end{array}
\right|\\
&=
\left|
\begin{array}{ccc}
1 & qVcos\phi_{n} \\
0 & 1
\end{array}
\right|\\
&=1
\end{aligned}
\end{equation}
Obviously, the mapping from $(\Delta E_{n}, \phi_{n})$ to $(\Delta E_{n+1}, \phi_{n+1})$ preserves the phase space area. The mapping algorithm is therefore stable. Because of the simplicity and stability of the mapping equations, they are frequently used in particle tracking calculations.
\section{Synchrotron Hamiltonian}
Although Eq.\ref{eq1} can track the evolution of particles numerically, it is more transparent to work with the Hamiltonian for a qualitative understanding of bunches of particles\cite{Weng1}. Here, we introduce the phase angle $\phi$ canonically conjugate to $\frac{\Delta E}{\omega_0}$, so that the phase space has the dimensions of action (energy times time), where $\omega_0$ is the angular revolution frequency of the on-momentum particle. In the conjugate phase space coordinates $(\phi,\frac{\Delta E}{\omega_0})$, the differential synchrotron equations of motion are
\begin{equation}
\label{eq4}
\left\{
\begin{aligned}
\frac{d}{dt}\left(\frac{\Delta E}{\omega_0}\right) & =  \frac{1}{2\pi}qV(sin\phi-sin\phi_s) \\
\frac{d}{dt}\phi & =  \frac{\omega_0^2 h\eta}{\beta^2E}\frac{\Delta E}{\omega_0}
\end{aligned}
\right.
\end{equation}
If the Hamiltonian doesn't contain time explicitly, or Hamiltonian is adiabatic, we can say
\begin{equation}
\label{eq5}
\frac{dH}{dt}=\frac{\partial H}{\partial t} \equiv 0
\end{equation}
So, Hamiltonian can be  a integral  as follows
\begin{equation}
\label{eq6}
H\left(\phi,\frac{\Delta E}{\omega_0}\right)=\int_{\phi_s}^\phi\frac{\partial H}{\partial \phi}d\phi+\int_{0}^{\frac{\Delta E}{\omega_0}}\frac{\partial H}{\partial \frac{\Delta E}{\omega_0}}d\left(\frac{\Delta E}{\omega_0}\right)
\end{equation}
Combining Eq.\ref{eq4} and the canonical equation
\begin{equation}
\label{eq7}
\left\{
\begin{aligned}
\frac{\partial H}{\partial \frac{\Delta E}{\omega_0}} & =  \frac{d}{dt}\phi\\
-\frac{\partial H}{\partial \phi} & =  \frac{d}{dt}\left(\frac{\Delta E}{\omega_0}\right)
\end{aligned}
\right.
\end{equation}
we obtain
\begin{equation}
\label{eq8}
\begin{aligned}
H\left(\phi,\frac{\Delta E}{\omega_0}\right)&=\frac{\omega_0^2h\eta}{2\beta^2E}\left(\frac{\Delta E}{\omega_0}\right)^2\\
&+\frac{qV}{2\pi}\left[cos\phi-cos\phi_s+(\phi-\phi_s)sin\phi_s\right]
\end{aligned}
\end{equation}
Through simple analysis, we can easily know that the Hamiltonian has two fixed point $(0,\phi_s)$ and $(0,\pi-\phi_s)$ corresponding to coordinates $(\frac{\Delta E}{\omega_0}, \phi)$. $(0,\phi_s)$ is the stable fixed point, or centre point, and $(0,\pi-\phi_s)$ is the unstable fixed point, or saddle point. The Hamiltonian torus passes through saddle point is called the separatrix, which defines the longitudinal phase space area, or bucket area. Substituting saddle point $(0,\pi-\phi_s)$ to Eq.\ref{eq8}, the Hamiltonian value of the separatrix can be written as
\begin{equation}
\label{eq9}
H_{sx}=\frac{qV}{2\pi}\left[-2cos\phi_s+(\pi-2\phi_s)sin\phi_s\right]
\end{equation}
Thus, the phase space trajectory of the separatrix is $H=H_{sx}$, or
\begin{equation}
\label{eq10}
\Delta E_{sx}^2+\frac{qV\beta^2E}{\pi h\eta}\left[cos\phi+cos\phi_s-(\pi-\phi-\phi_s)sin\phi_s\right]=0
\end{equation}
Bucket height or maximum energy deviation is
\begin{equation}
\label{eq11}
\begin{aligned}
\Delta E_{max}&=\Delta E_{sx}(\phi_s)\\
&=\sqrt{-\frac{qV\beta^2E}{\pi h\eta}\left[2cos\phi_s-(\pi-2\phi_s)sin\phi_s\right]}
\end{aligned}
\end{equation}
Bucket area, namely the phase space area enclosed by the separatrix, is
\begin{equation}
\label{eq12}
\begin{aligned}
A_B&=\iint d\Delta E_{sx}d\phi=\oint \Delta E_{sx}(\phi)d\phi\\
&=16\sqrt{\frac{qV\beta^2E}{2\pi h|\eta|}}\alpha_b(\phi_s),\alpha_b(\phi_s)\approx \frac{1-sin\phi_s}{1+sin\phi_s}
\end{aligned}
\end{equation}
In the process of capture, $\phi_s=0$, so
\begin{equation}
\label{eq13}
\Delta E_{max}=2\sqrt{\frac{qV\beta^2E}{2\pi h|\eta|}}
\end{equation}
and
\begin{equation}
\label{eq14}
A_B=16\sqrt{\frac{qV\beta^2E}{2\pi h|\eta|}}
\end{equation}
Obviously, the phase space area and bucket height for a stationary bucket have the following relation,
\begin{equation}
\label{eq15}
A_B=8\Delta E_{max}
\end{equation}
\section{Adiabatic capture voltage curve}
\begin{figure}[!htb]
\includegraphics[width=8cm]{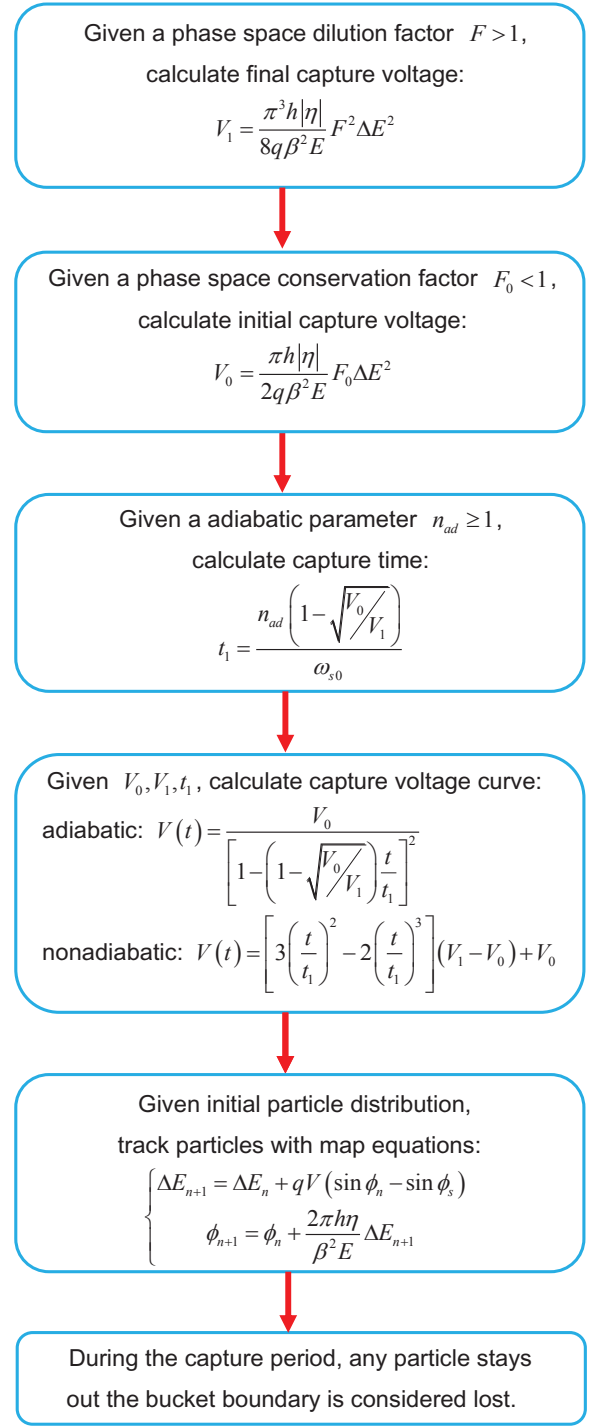}
\caption{\label{fig2}   Simulation flow chart. }
\end{figure}
To capture coasting beam (it is true for most sychrotrons) to a bucket without loss, the bucket area after capture finish should be chosen so that it is large enough compared to initial beam area. Assume that the energy deviation of the injected beam is $\Delta E$, the phase space area of the coasting beam in one RF period is
\begin{equation}
\label{eq16}
A_0=4\pi\Delta E
\end{equation}
Let $A_B=FA_0$, in which $F>1$ represents the phase space dilution factor. Thus, final capture voltage can be calculated via $A_B=FA_0$, that is
\begin{equation}
\label{eq17}
V_1=\frac{\pi^3h|\eta|}{8q\beta^2E}F^2\Delta E^2
\end{equation}
Usually, we select $F=\frac{3}{2}$\cite{LiuWei}.

To preserve emittance as a constant as much as possible in the process of capture, according to numerous experience, the initial bucket area should be much smaller than the coasting beam area\cite{NG1}. Suppose the bucket height equals initial energy deviation, that is $\Delta E_{max}=\Delta E$. Hence, the corresponding voltage is
\begin{equation}
\label{eq18}
V=\frac{\pi h|\eta|}{2q\beta^2E}\Delta E^2.
\end{equation}
So, initial capture voltage should obey $V_0<V$. Let $V_0=F_0V$, in which $F_0<1$ represents the phase space conservation factor. Thus, the initial capture voltage is
\begin{equation}
\label{eq19}
V_0=\frac{\pi h|\eta|}{2q\beta^2E}F_0\Delta E^2
\end{equation}
In principle, the smaller the $F_0$, the more conservation the emittance, however, much smaller $F_0$ leads to a much longer capture time. Therefore, $F_0$ should be reasonable in some sense, i.e. 0.2.

Adiabatic capture criterion\cite{Ng2} states that the relative change in bucket height or bucket area should be much slower than the synchrotron frequency $\omega_s/2\pi$, or
\begin{equation}
\label{eq20}
\omega_s\gg \frac{1}{A}\frac{dA}{dt}
\end{equation}
Since bucket area $A$ and synchrotron angular frequency $\omega_s$ are both proportional $\sqrt{V}$, so we can write down
\begin{equation}
\label{eq21}
A=k_1\sqrt{V}, \omega_s=k_2\sqrt{V}
\end{equation}
Substituting Eq. \ref{eq21} to Eq. \ref{eq20}, we have
\begin{equation}
\label{eq22}
\frac{1}{2V^{3/2}}\frac{dV}{dt}\ll k
\end{equation}
After integral,
\begin{equation}
\label{eq23}
V(0)^{-1/2}-V(t)^{-1/2}\ll kt
\end{equation}
Let $V(0)=V_0$, and then initial synchrotron angular frequency can be denoted as $\omega_{s0}=k\sqrt{V_0}$. So, above equation can be written as
\begin{equation}
\label{eq24}
\frac{\omega_{s0}t}{1-\sqrt{V_0/V(t)}}\gg 1
\end{equation}
Defining adiabatic parameter $n_{ad}=\frac{\omega_{s0}t}{1-\sqrt{V_0/V(t)}}$, thus adiabatic condition becomes $n_{ad}\gg 1$. Hence, adiabatic capture voltage is
\begin{equation}
\label{eq25}
V(t)=\frac{V_0}{\left(1-\omega_{s0}t/n_{ad}\right)^2}
\end{equation}
Assume the capture time and voltage change from $(V_0,0)$ to $(V_1,t_1)$ in the process of adiabatic capture,
where $V_0,V_1,t_1$ are initial capture voltage, final capture voltage and capture time respectively, so
\begin{equation}
\label{eq26}
V_1=\frac{V_0}{\left(1-\omega_{s0}t_1/n_{ad}\right)^2}
\end{equation}
Substituting Eq. \ref{eq26} to Eq. \ref{eq25}, finally, RF capture voltage curve can be formulated as
\begin{equation}
\label{eq27}
V(t)=\frac{V_0}{\left[1-\left(1-\sqrt{V_0/V_1}\right)\frac{t}{t_1}\right]^2}
\end{equation}
If $V_0,V_1,n_{ad}$ are given, capture time can be easily calculated as follows,
\begin{equation}
\label{eq28}
t_1=\frac{n_{ad}\left(1-\sqrt{V_0/V_1}\right)}{\omega_{s0}}
\end{equation}
where $\omega_{s0}=\omega_{0}\sqrt{\frac{h|\eta|qV_0}{2\pi\beta^2E}}$, and $\omega_{0}$ is the angular revolution frequency.
\section{Longitudinal RF capture simulation}
\begin{figure*}
\centering
\includegraphics[width=16cm,height=21cm]{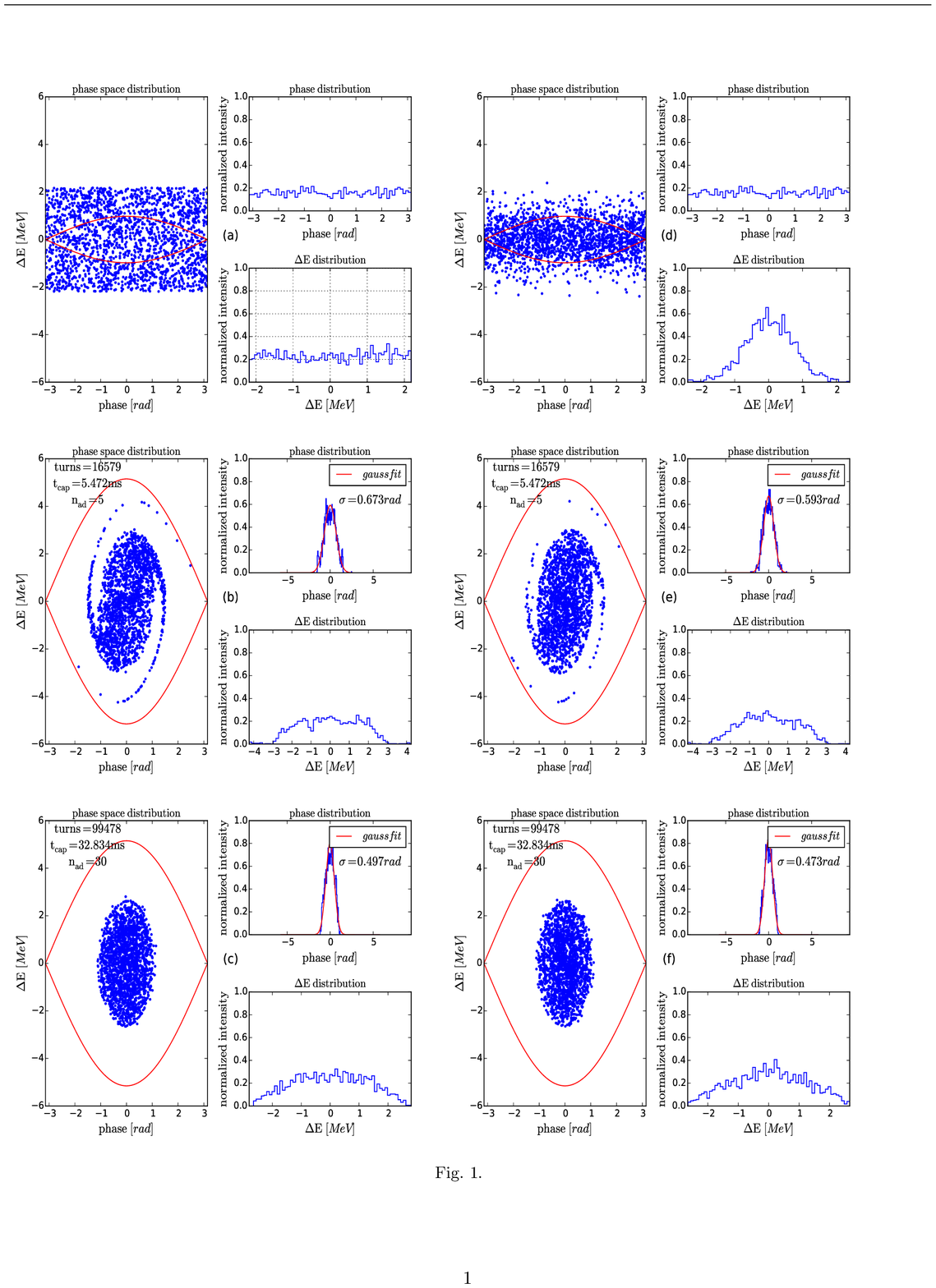}
\caption{\label{fig3}Plot of longitudinal phase space evolution after adiabatic capture with different adiabaticity under adiabatic voltage curve. The initial and final capture voltage are fixed at $19.9V$ and $553V$ and the initial bunch area is $27.494 MeVrad$. (a) is coasting beam with uniform distribution and energy deviation is $2.188MeV$. (b) is adiabatic capture with adiabatic parameter $n_{ad}=5$ corresponding to capture time $5.472ms$ or $16579$ turns. (c) is adiabatic capture with adiabatic parameter $n_{ad}=30$ corresponding to capture time $32.834ms$ or $99478$ turns. (d) is coasting beam with Gaussian distribution and RMS energy deviation is $0.729MeV$. (e) and (f) are adiabatic capture with the same settings with (b) and (c) respectively. }
\end{figure*}
\begin{figure*}
\centering
\includegraphics[width=16cm,height=21cm]{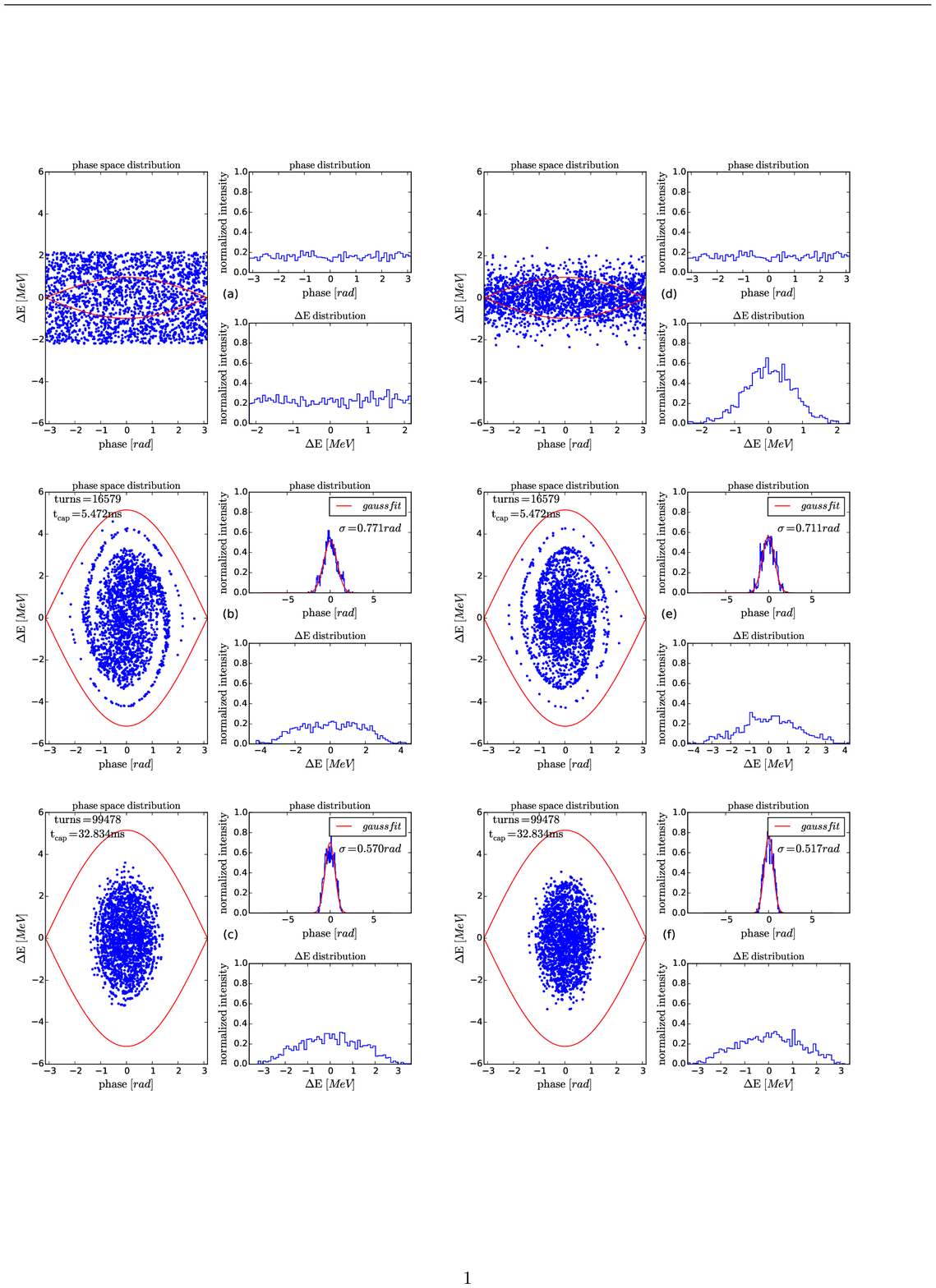}
\caption{\label{fig4} Plot of longitudinal phase space evolution after nonadiabatic capture under Kang's voltage curve. The initial and final capture voltage are fixed at $19.9V$ and $553V$ and the initial bunch area is $27.494 MeVrad$. (a) is coasting beam with uniform distribution and energy deviation is $2.188MeV$. (b) is nonadiabatic capture with capture time $5.472ms$ or $16579$ turns. (c) is nonadiabatic capture with capture time $32.834ms$ or $99478$ turns. (d) is coasting beam with Gaussian distribution and RMS energy deviation is $0.729MeV$. (e) and (f) are nonadiabatic capture with the same settings with (b) and (c) respectively. }
\end{figure*}

The simulation flow chart is illustrated in Fig. \ref{fig2}. Initial conditions used in this simulation: particle $^{12}C^{6+}$, kinetic energy $E_k=200MeV/u$, fractional momentum deviation $\frac{\Delta p}{p}=\pm 5.0\times 10^{-4}$ or energy deviation $2.188MeV$, harmonic number $1$, transition energy $1.741$, phase space dilution factor $F=1.5$, phase space conservation factor $F_0=0.2$, the number of macro-particles $2000$. The initial coasting beam area is $A_0=4\pi\Delta E=4\pi\times 2.188 MeVrad=27.494MeVrad$. Considering phase space conservation factor and phase space dilution factor, the initial capture voltage and final capture voltage are $19.9V$ and $553V$, and the corresponding bucket area or bucket height is $7.827MeVrad$ or $0.978MeV$ and $41.241MeVrad$ or $5.155MeV$, respectively. The initial bucket area is small enough with respect to initial coasting beam area, and meanwhile, the final bucket area is large enough with respect to initial coasting beam area, which guarantees high capture efficiency.

Simulation of adiabatic capture with different adiabatic parameters is illustrated in Fig.\ref{fig3}. In the case of uniform distribution, for $n_{ad}=5$, the corresponding capture time is $5.472ms$ or $16579$ turns, and the final bunch length is $0.673rad$, which is large compared to the case of $n_{ad}=30$, i.e. $0.497rad$. And meanwhile, the phase space area is also large. So, small adiabatic parameter leads to short capture time, and therefore the bad capture. In addition, it is apparent that Gaussian distribution performs a better capture compared to uniform distribution, because Gaussian distribution has a dense beam core.

\begin{figure}[!htb]
\includegraphics[width=8cm]{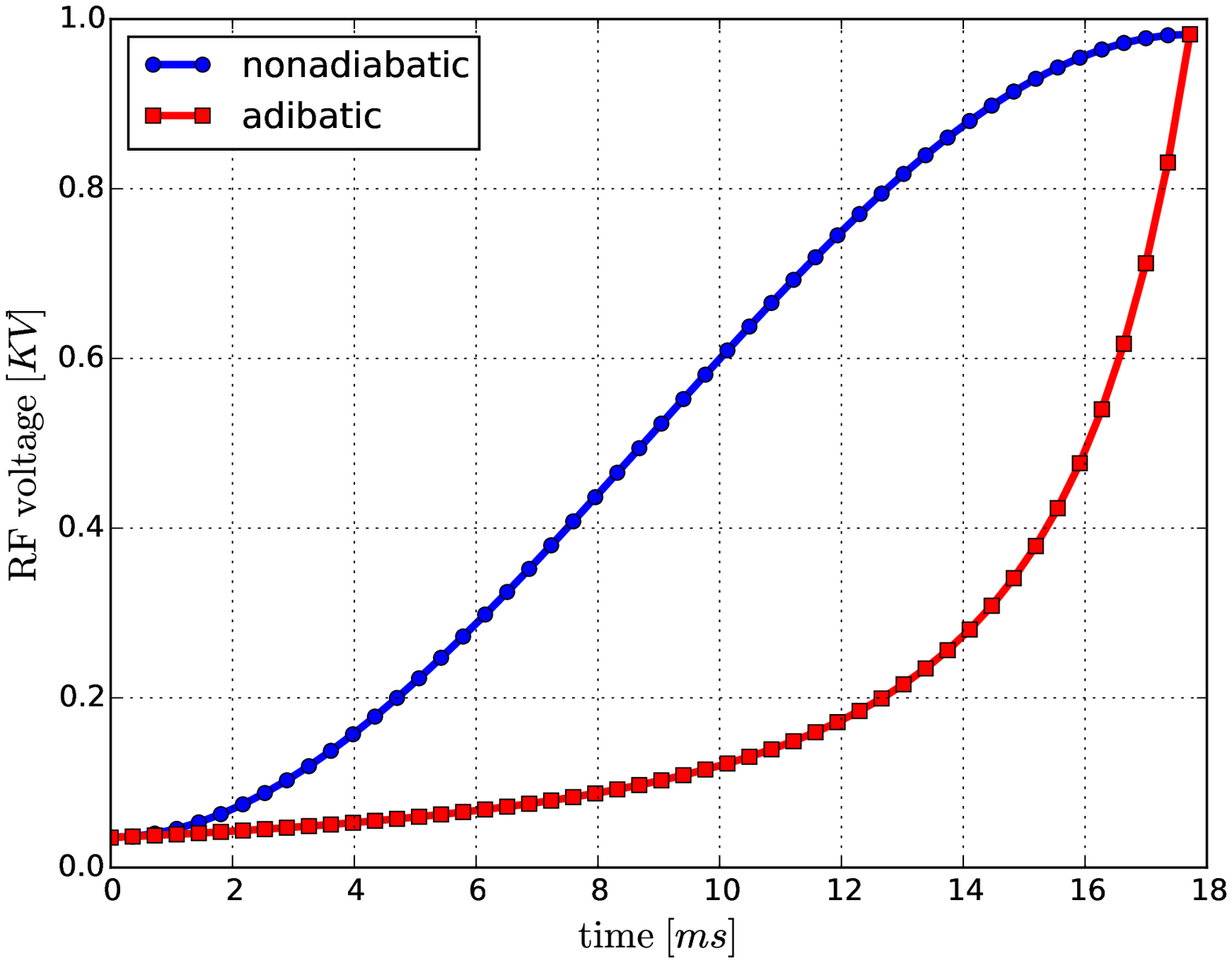}
\caption{\label{fig5}   Two kinds of capture voltage curves. }
\end{figure}

As a comparison, we introduce another kind of capture voltage curve, namely Kang's curve or nonadiabatic curve, which is employed by Kang Xiaojian to study adiabatic capture of a linac beam into buckets of the CIS booster of the IUCF Electron Cooler\cite{Kang}. The RF voltage used has the following form
\begin{equation}
\label{eq29}
V(t)=\left[3\left(\frac{t}{t_1}\right)^2-2\left(\frac{t}{t_1}\right)^3\right](V_1-V_0)+V_0
\end{equation}
The simulation result for nonadiabatic capture is shown in Fig.\ref{fig4}. The simulation parameters applied are identical to adiabatic capture, and also the phase space evolution is similar to adiabatic situation. However, the capture is much worse under the identical capture time compared to adiabatic capture, which performs a more loose phase space distribution and longer bunch length. The reason why this happens is that the much more rapid increase in RF voltage in nonadiabatic curve leads to a heavy filament of bunch at the beginning of the capture, which is shown in Fig.\ref{fig5}. Therefore, slow increase of RF voltage, especially, in the beginning of capture, is essential for adiabaticity.

\begin{figure}[!htb]
\includegraphics[width=8cm]{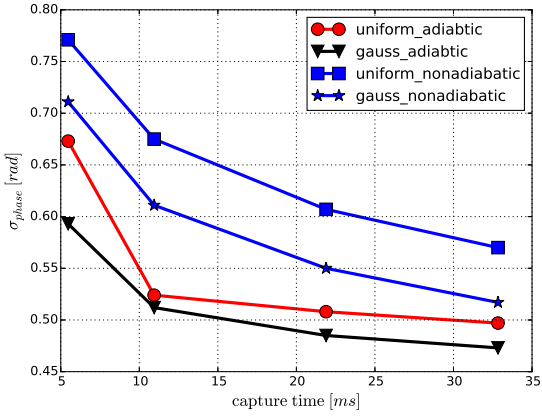}
\caption{\label{fig6}   Bunch length or phase distribution with capture time. }
\end{figure}

From Fig.\ref{fig6}, we can see that, with the increase of capture time, the capture becomes more and more better. And with very long capture time or extremely slow capture, the four kinds of results become roughly similar, which indicates that the capture process reaches the adiabatic limit for either one case. Careful analysis reveals that Gaussian distribution with adiabatic capture is the best one. Meanwhile, to reach similar capture result, adiabatic capture takes less time than nonadiabatic capture. Finally, whatever which one case, there is no beam loss, because the final bucket area is large enough and the initial energy deviation is small enough.
\begin{figure}[!htb]
\includegraphics[width=8cm]{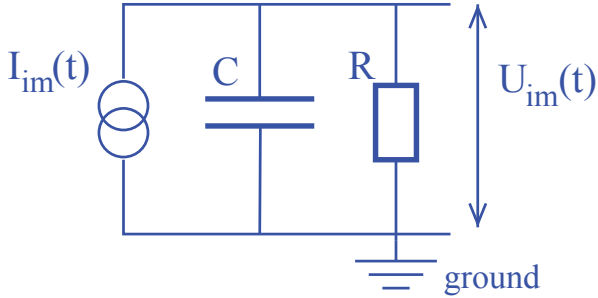}
\caption{\label{fig7}   Equivalent circuit for a typical capacitive BPM. }
\end{figure}
\section{BPM signal estimation}
As described in introduction, BPM is a non-destructive diagnostics tool used most frequently at linacs, transfer line, and synchrotrons. The equivalent circuit\cite{Forck2011} for a typical capacitive BPM is shown in Fig.\ref{fig7}. Where $I_{im}(t)$ is the image current driven by the image charge on the plates, $U_{im}(t)$ is the induced voltage that is the signal we can observe directly, $C$  is the equivalent capacity of the BPM electrode against the beam pipe, and $R$ is the input resistor of amplifier.

For such a circuit, the impedance can be written easily as
\begin{equation}
\label{eq30}
Z=\frac{R}{1+j\omega RC}
\end{equation}
Hence, the voltage drop at the impedance $Z$ is
\begin{equation}
\label{eq31}
U_{im}(t)=ZI_{im}(t)
\end{equation}
The image current can be expressed as a function of beam current,
\begin{equation}
\label{eq32}
\begin{aligned}
I_{im}(t)\equiv \frac{dQ_{im}(t)}{dt}
&=k\frac{dQ_{beam}(t)}{dt}\\
&=K \frac{dI_{beam}(t)}{dt}\\
&=K j\omega I_{beam}(t)
\end{aligned}
\end{equation}
in which, $k$ and $K$ are constant, $I_{beam}(t)=I_0e^{j\omega t}$. So, time domain signal is
\begin{equation}
\label{eq33}
U_{im}(t)=K \frac{j\omega RC}{1+j\omega RC}I_{beam}(t)=Z(\omega)I_{beam}(t)
\end{equation}
Because $Z(\omega)$ is a frequency domain value, we have to transform $I_{beam}(t)$ to frequency domain, i.e. via FFT, which yields
\begin{equation}
\label{eq34}
U_{im}(\omega)=Z(\omega)I_{beam}(\omega)
\end{equation}
After inverse FFT, the time domain voltage signal becomes back to $U_{im}(t)$.

In the high frequency range, or $f\gg f_{cut}$, we have
\begin{equation}
\label{eq35}
Z(\omega)\propto \frac{j\omega/\omega_{cut}}{1+j\omega/\omega_{cut}}\approx 1
\end{equation}
Thus,
\begin{equation}
\label{eq36}
U_{im}(t)\propto I_{beam}(t)
\end{equation}
Therefore, voltage signal is the direct image of beam current in the range of high frequency.
However, in the low frequency range, or $f\ll f_{cut}$, we have
\begin{equation}
\label{eq37}
Z(\omega)\propto \frac{j\omega/\omega_{cut}}{1+j\omega/\omega_{cut}}\approx j\frac{\omega}{\omega_{cut}}
\end{equation}
Thus,
\begin{equation}
\label{eq38}
U_{im}(t)\propto \frac{dI_{beam}(t)}{dt}
\end{equation}
Therefore, voltage signal is the derivative of beam current in the range of low frequency.
\begin{figure*}
\centering
\includegraphics[width=15cm]{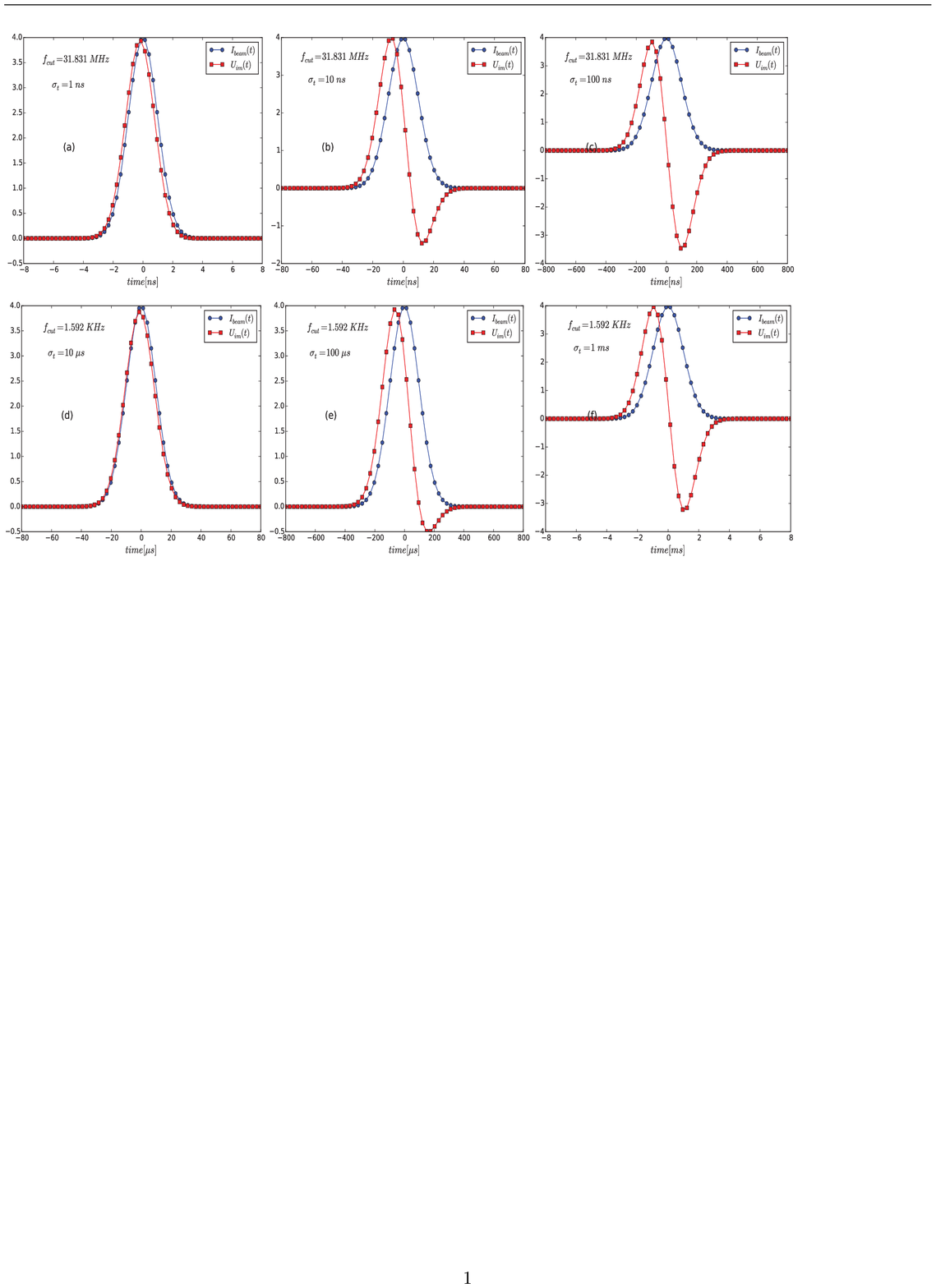}
\caption{\label{fig8}Simulated BPM signal for various bunch length with high and low input impedance. Figure (a), (b) and (c) correspond to input impedance of $50\Omega$, and bunch length is $1ns$, $10ns$ and $100ns$, respectively. Figure (d), (e) and (f) correspond to input impedance of $1\Omega$, and bunch length is $10\mu s$, $100\mu s$ and $1ms$, respectively. The solid blue curve with circle is the beam current distribution. }
\end{figure*}

The cut-off frequency for Fig.\ref{fig7} described circuit is $f_{cut}=\frac{1}{2\pi RC}$. For the case of so-called linear cut BPM or shoe box BPM used most often at synchrotrons, the typical value of the capacitance is about $C\approx 100pF$. Therefore, the cut-off frequency for low input impedance, i.e. $50\Omega$, is $f_{cut}=\frac{1}{2\pi\times 50 \times 10^{-10}}=31.831MHz$, and for high input impedance, i.e. $1M\Omega$, is $f_{cut}=\frac{1}{2\pi\times 10^6 \times 10^{-10}}=1.592KHz$. For a Gaussian current distribution in time domain with a width of $\sigma_t$, after FFT, it is still a Gaussian distribution, however, with a width of $\sigma_f=\frac{1}{2\pi\sigma_t}$ instead. In the case of low input impedance, if the bunch length is $\sigma_t=1ns$, and hence $f=159.155MHz$, the BPM voltage signal is a direct image of beam structure since the frequency is much larger than cut-off frequency. The calculated signal shape is shown in Fig.\ref{fig8}(a). If the bunch length is $\sigma_t=100ns$ or frequency $f=1.592MHz$, the signal is a derivative of beam structure since the frequency is much smaller than cut-off frequency, as shown in Fig.\ref{fig8}(c). And if the bunch length is $\sigma_t=10ns$ or frequency $f=15.915MHz$, the signal is a intermediate state since the frequency is comparable to cut-off frequency, which is shown in Fig.\ref{fig8}(b). For the case of longer bunch, the simulation results are shown in Fig.\ref{fig8}(d),(e) and (f), which is corresponding to high impedance case, i.e. $1M\Omega$ or $f_{cut}=1.592KHz$, and similar results are achieved. According to the simulation, we can conclude that the signal is always a derivative of beam current for a very long bunch, i.e. synchrotrons, and a direct image of beam current for a very short bunch, i.e. electron machine.
In the previous capture simulation, the smallest bunch length is $\sigma_{\phi}=0.473rad$ or $\sigma_t=\frac{\sigma_{\phi}}{2\pi f_0}=\frac{0.473}{2\pi\times 10^6}=75.3ns$ corresponding to revolution frequency $1MHz$, which is similar to the case of $\sigma_t=100ns$ with input impedance of $50\Omega$. The longest bunch length is $\sigma_{\phi}=0.771rad$ or $\sigma_t=123ns$, which is also similar to the case of $\sigma_t=100ns$. However, if the input impedance is $1M\Omega$, the signal is the direct image of beam current, which is similar to the case of $\sigma_t=10\mu s$. For HIMM, the BPM has a low input impedance $50\Omega$, so the signal should be a derivative of beam current. In the next section we will present the BPM signal measured at HIMM.

\section{HIMM BPM signal}
\begin{figure}
\includegraphics[width=8cm]{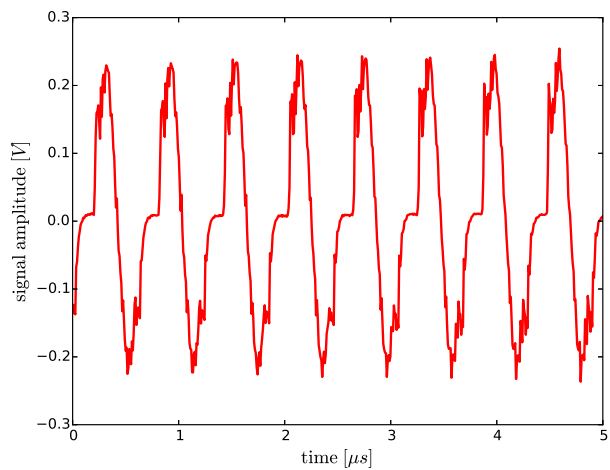}
\caption{\label{fig9}   Measured BPM signal of HIMM. }
\end{figure}

Heavy Ion Medical Machine (HIMM)\cite{Yang} is a new compact accelerator facility dedicated to carbon cancer therapy, which having been constructed in Wuwei and Lanzhou city. Eight BPMs are installed along the synchrotron, which are mainly used to control beam orbit.
The BPM is linear cut type with a low input impedance $50\Omega$, and the capacity is $100pF$. The raw analog signal is sent to NI 5734 sample card that has a sample rate of $120MS/s$ with 16 bit, and a attenuator is adapted to avoid signal saturation. A Labview based program is developed to acquire turn-by-turn data. For such machine, the bunch length is about $100ns$ according to simulation, so the signal shape is the same as Fig.\ref{fig8}(c). The measured BPM signal is shown in Fig.\ref{fig9}. Obviously, an expected agreement is presented.
\section{Conclusion}
In this paper, the longitudinal RF capture theory is reviewed in details and a simulation code is developed. The simulation of capture with various distribution and capture time is performed. A comparison between adiabatic and nonadiabatic capture is discussed, and adiabatic capture with Gaussian distribution is the best choice for capture process. Meanwhile, we can conclude that capture time is essential for the whole capture process and the longer the capture time the better the capture. In addition, BPM signal with different input impedance is simulated and compared to measured one from HIMM, and a good agreement is achieved. Hence, this method to simulate BPM signal is absolutely feasible.

\section*{References}

\bibliography{nima}

\end{document}